\documentclass[10pt]{article}
\usepackage{graphicx}
\begin{document}

\begin{center}
{\bf \large Physical parameters of components in close binary systems: V} 
\end{center}

\begin{center}
by
\end{center}

\begin{center}
\large S.~Zola$^{1,2}$, J.M.~Kreiner$^{2}$, B.~Zakrzewski$^{2}$, D.P.~Kjurkchieva$^{3}$,
       D.V.~Marchev$^{3}$, A.~Baran$^{2,4}$,  S.M.~Rucinski$^{5}$, W.~Ogloza$^{2}$,   
       M.~Siwak$^{1}$, D.~Koziel$^{1}$, M.~Drozdz$^{2}$ and B.~Pokrzywka$^{2}$
\end{center}

\begin{center}
$^1$ Astronomical Observatory of the Jagiellonian University,
     ul. Orla 171, 30-244 Cracow, Poland \\
     email: siwak@oa.uj.edu.pl \\

$^2$ Mt. Suhora Observatory of the Pedagogical University,
     ul. Podchor\c{a}\.{z}ych 2, 30-084 Cracow, Poland \\
     email: sfkreine@cyf-kr.edu.pl\\

$^3$ Shumen University, 115 Universitetska Str, 9700 Shumen,\\
       Bulgaria\\

$^4$ Torun Centre for Astronomy, N. Copernicus University, \\
     ul. Gagarina 11, 87-100 Torun, Poland \\

$^5$ David Dunlap Observatory, University of Toronto,
     P.O. Box 360, Richmond Hill, Ontario, Canada L4C 4Y6 \\
     email: rucinski@astro.utoronto.ca \\

\end{center}

\begin{center}
ABSTRACT
\end{center}

The paper presents combined spectroscopic and photometric orbital
solutions for ten close binary systems: CN And, V776 Cas, FU Dra,
UV Lyn, BB Peg, V592 Per, OU Ser, EQ Tau, HN UMa and HT Vir. The
photometric data consist of new multicolor light curves,
while the spectroscopy has been recently obtained within the
radial velocity programme at the David Dunlap Observatory (DDO).
Absolute parameters of the components for these binary systems are
derived. Our results confirm that CN And is not a contact system.
Its configuration is semi-detached with the secondary component filling
its Roche lobe. The configuration of nine other systems is
contact. Three systems (V776 Cas, V592 Per and OU Ser) have high (44-77\%)  
and six (FU Dra, UV Lyn, BB Peg, EQ Tau, HN UMa and HT Vir) low or
intermediate (8-32\%) fill-out factors. The absolute physical parameters
are derived. \\

{\noindent}

\begin{center}
{\bf 1. ~Introduction}
\end{center}

This paper is a continuation of a series of papers (Kreiner et al.
2003 (Paper I), Baran et al. 2004 (Paper II), Zola et al. 2004 (Paper III), 
Gazeas et al. 2005 (Paper IV)
aimed at derivation of the physical parameters of the components in contact 
and close binary systems with the ultimate goal of rediscussing the properties 
of components (including their evolutionary status)  based on a sample 
of few dozen such binaries, the parameters of which were determined 
in a uniform way.
As has been shown (see DDO series of papers - Rucinski et al. 2005 and references 
therein),  determination of absolute 
parameters of contact systems only from light curve modeling, without knowing the 
spectroscopic mass ratio, can often result in spurious solutions thus leading 
to unreliable physical parameters of components. 

The solutions based on recently obtained spectroscopic
and multicolor photometric data for the next ten binary stars: CN And, V776
Cas, FU Dra, UV Lyn, BB Peg, V592 Per, OU Ser, EQ Tau,
HN UMa and HT Vir are described in sections 2-4. The results derived are 
briefly discussed in Section 5.

\begin{center}
{\bf 2. ~The targets}
\end{center}

{\it 2.1. ~CN And} \\

CN And (BD+39$^{\circ}$0059, GSC~02787-01815, V=9.76$^{m}$) was 
discovered by Hoffmeister (1949) as a short-period eclipsing
binary. 
Photometric observations of CN And have been carried out by
Bozkurt et al. (1976), Seeds and Abernethy (1982), Michaels et
al. (1984), Rafert et al. (1985), Keskin (1989), Samec et al.
(1998).
The orbital period and light curve changes have been investigated by 
Evren et al. (1987) and Keskin (1989).
Samec et al. (1998)  interpreted the decreasing orbital period 
during the past 50 yr as being due to the mass transfer from the primary 
to the secondary component and/or caused by magnetic breaking as a result 
of strong system activity. Van Hamme et al.
(2001) determined the rate of the mass transfer as 1.4 x 10$^{-7}$
M$_{\odot}$/yr.

CN And is an active solar type binary with components of spectral
type in the F5 to G5 range. Flare events with an amplitude of 0.11
mag at phases 0.21-0.24 (Yang and Liu 1985) and moderately
strong X-ray emission (Shaw et al. 1996) from this star have been
detected.
Rucinski et al. (2000) obtained radial velocity curves of both components and  
determined 
the following spectroscopic elements of the orbit: $\gamma$=-24.9
km/s; K$_{1}$=87.5 km/s; K$_{2}$=224.3 km/s;
(M$_{1}$+M$_{2}$)sin$^{3}$i=1.46 M$_{\odot}$; q=0.39. This value of
mass ratio differs considerably from the photometric estimates ( 0.5$<$q$<$0.8) 
of Kaluzny (1983) and Rafert et al. (1985). Rucinski et al. (2000) concluded 
that CN And looks either like a typical contact binary of A sub-type or
it could be a very close semidetached binary.  
Recently, \c{C}i\c{c}ek, et al. (2005) obtained new BVR light curves and computed
absolute elements of the components. They concluded that CN And is almost a contact
binary and the components fill about 99$\%$ of their Roche lobes.

According to Kaluzny (1983) CN And is a promising candidate for
identification as a W UMa-type system caught in a phase of poor
thermal contact, a phase  predicted by the
thermal relaxation oscillation theory (Lucy and Wilson 1979).\\

\begin{table}
\begin{center}
\caption []{Journal of photometric observations}
%\begin{flushleft}
\begin{tabular}{lll}
\hline
 object  & observatory & dates             \\
\hline
 CN And  & Mt. Suhora       &  19/20, 25/26, 29/30 Aug, 31~Aug/1~Sep, 3/4 Sep 2002\\
 V776 Cas& Mt. Suhora       &  19/20, 24/25 Oct 2005    \\
 FU Dra  & Jagiell. Univ.   &  27/28 Feb, 1/2, 4/5, 13/14 Mar 2002    \\
 UV Lyn  & Mt. Suhora       &  5/6, 6/7 Feb 2005   \\
 BB Peg  & Mt. Suhora       &  10/11 Aug, 5/6, 6/7 Oct 2004 \\
 V592 Per& Mt. Suhora       &  2/3, 4/5, 9/10, 10/11 Dec 2004, 16/17, 28/29 Jan 2005    \\
 OU Ser  & Mt. Suhora       &  4/5, 5/6, 6/7 May 2003   \\
 EQ Tau  & Mt. Suhora       &  15/16 Dec 2004, 8/9,  9/10 Jan 2005 \\
 HN UMa  & Mt. Suhora       &  27/28 Feb 2003 \\
 HT Vir  & Mt. Suhora       &  18/19, 19/20, 20/21, 21/22, 22/23, 23/24, April  14/15, 20/21 May 2004     \\
\hline
\end{tabular}
%\end{flushleft}
\end{center}
\label{Tab1}
\end{table}

{\it 2.2. ~V776 Cas} \\

V776 Cas (BD+69$^{\circ}$0121, HIP 8821, V=9.00$^{m}$) is the brighter 
member of the visual binary
ADS 1485. The companion, at separation of 5.38 arcsec, is 2 mag
fainter than the contact binary.
V776 Cas was discovered as an eclipsing binary with a small
amplitude by the Hipparcos mission (ESA, 1997). It appeared in Duerbeck (1997)
as V778 Cas with twice the actual period,
spectral type F0, B-V=0.525 and variability range of 8.94-9.09
mag in V.
The star was observed photometrically by Gomez-Forrellad et al.
(1999) who assumed the star to be  an EW-type binary undergoing
marginal eclipses or an ellipsoidal variable. Their V light curve
showed the primary minimum being deeper than the secondary by 0.04 mag,
two almost equal in height maxima  and the amplitude of variability
of about 0.19 mag. 
Rucinski et al. (2001) obtained the first spectroscopic orbit of
V776 Cas: $\gamma$=-24.7 km/s; K$_{1}$=32 km/s; K$_{2}$=245.3
km/s; q=m$_{c}$/m$_{h}$=0.13; (M$_{1}$+M$_{2}$)sin$^{3}$i=0.975
M$_{\odot}$. They found that the spectral type of the system is F2V
and that it belongs to the A sub-class of the contact binaries.
A photometric study of this system was published by Djura\v{s}evi\'c et 
al. (2004).\\

{\it 2.3. ~FU Dra} \\

The contact binary FU Dra (HIP 076272, GSC~04181-00673, V=10.55$^{m}$) was discovered 
by the Hipparcos mission.
The system  has a large proper motion.
Rucinski et al. (2000) found that FU Dra belongs to the W-type
subgroup with spectral type F8V and obtained the following
spectroscopic elements of the orbit of FU Dra: $\gamma$=-11.4
km/s; K$_{1}$=70.4 km/s; K$_{2}$=280.8 km/s; q=0.251;
(M$_{1}$+M$_{2}$)sin$^{3}$i=1.38 M$_{\odot}$.  
Va\v{n}ko et al. (2001)  obtained new photoelectric and CCD observations and 
determined the absolute parameters of the system. \\

\begin{table}
\begin{center}
\caption{Linear ephemerides used for phase calculation and comparison stars 
         used in our observations}
%\begin{flushleft}
\begin{tabular}{lcll}
\hline
 star    & reference epoch (HJD) & period (days)  & comparison star \\
\hline
 CN And  & 2452500.1210          & 0.4627899      & GSC~02786-01556 \\
 V776 Cas& 2452932.4339          & 0.44041574     & GSC~04314-00449 \\
 FU Dra  & 2452338.5183          & 0.3067171      & GSC~04181-01726 \\
 UV Lyn  & 2453407.3606          & 0.41498402     & GSC~02983-01770 \\
 BB Peg  & 2453228.4574          & 0.361501       & GSC~01682-01444 \\
 V592 Per& 2453399.3400          & 0.715722       & GSC~02897-00882 \\
 OU Ser  & 2452500.0710          & 0.296759       & GSC~01487-00941 \\
 EQ Tau  & 2453354.5303          & 0.341348       & GSC~01800-02137 \\
 HN UMa  & 2452698.3008          & 0.382525367    & GSC~02522-00553 \\
 HT Vir  & 2453117.4702          & 0.407671       & GSC~00311-01613 \\
\hline
\end{tabular}
%\end{flushleft}
\end{center}
\label{Tab2}
\end{table}

{\it 2.4. ~UV Lyn} \\

UV Lyn (BD+38$^{\circ}$1992, HIP 44455, V=9.58$^{m}$)  was discovered to be variable 
by Kippenhahn (Geier et al. 1955). Kuklin (1961) and Strohmeier et al. (1964) found 
1.2 day periodicity, but later Strohmeier suspected the period to be
incorrect. Bossen (1973) classified UV Lyn as a W UMa type binary
with a period of 0.415 day and maxima of unequal brightness.
On the basis of the light curve analysis Markworth and Michaels (1982) found that 
the star is a contact system with an inclination i=68$^{\circ}$ and a mass 
ratio q=0.526. Zhang et al. (1995) found a slow increase of the orbital period
(confirmed later by Va\v{n}ko et al. 2001). They explained the change
of the period by  mass transfer from the secondary to the primary
component. Lu and Rucinski (1999) obtained the first spectroscopic orbit 
of UV Lyn: $\gamma$=-0.3 km/s; K$_{1}$=86.5 km/s; K$_{2}$=235.7 km/s;
q=0.367; (M$_{1}$+M$_{2}$)sin$^{3}$i=1.44 M$_{\odot}$. \\

{\it 2.5. ~BB Peg} \\

BB Peg (HIP 110493, GSC~01682-01542, V=11.6$^{m}$) was discovered by Hoffmeister (1931).
Its period was revised by Whitney (1959) to be 0.3515 day. The BV
observations of Cerruti-Sola et al. (1981) revealed a variable
degree of asymmetry from the yellow to the blue light curve and a
phase shift of the secondary minimum. They classified BB Peg as
a contact system of the W sub-type with the fill-out parameter
of 37$\%$. Leung et al. (1985) solved their light curves and derived a
photometric mass ratio of 0.356 and the fill-out factor of 12$\%$. The UBV data of
Awadalla (1988) showed variable depth of the primary eclipse that is
a total occultation and an increase of the orbital period.
These features were explained by the existence of dark spots on the
secondary star near the L$_{1}$ point.
Hrivnak (1990) obtained the first spectroscopic value of the mass ratio q=0.34,
later re-determined to be q=0.36 by Lu and Rucinski (1999). Other orbital elements
were: $\gamma$=29.9 km/s; K$_{1}$=59.5 km/s; K$_{2}$=263.3
km/s; (M$_{1}$+M$_{2}$)sin$^{3}$i=1.86 M$_{\odot}$. \\

{\it 2.6. ~V592 Per} \\

The variability of V592 Per (BD+39$^{\circ}$1054, HIP 022050, V=8.37$^{m}$) was 
discovered  by Hipparcos. The star has been known to posses a near visual companion 
(Heintz 1990). Grenier et al. (1999) estimated the spectral type as F2IV,  
in agreement with the value (B-V)=0.35 from the Tycho-2 data (Hog et al. 2000).
Douglas et al. (2000) and Prieur et al. (2001) obtained specle observations
of the triple system. 
Rucinski et al. (2005) re-discovered the visual companion
spectroscopically. Although it is fainter than the close binary,
its spectral features are well defined due to their sharpness.
The contribution of the companion to the total light of the system
was estimated to be about 40$\%$. 
Rucinski et al. (2005) obtained the first spectroscopic orbit of
V592 Per: $\gamma$=27.9 km/s; K$_{1}$=93.7 km/s; K$_{2}$=230
km/s; q=0.408; (M$_{1}$+M$_{2}$)sin$^{3}$i=2.52 M$_{\odot}$. The
new spectroscopic mass ratio value  differs considerably from
that (q=0.25) derived from the light curve modeling by Selam (2004).  \\

\begin{table}
\begin{center}
\caption{Search ranges of the adjusted parameters}
\begin{tabular}{lcccc}
\hline
System & $i$(deg)&  $T_{\rm 2}({\rm K})$  & $\Omega_{\rm 1}$=$\Omega_{\rm 2}$ & $L_{1}$         \\
\hline
CN And  & 60-90 & 4800-6400     & 2.00-4.00     & 6.0-13.5       \\
V776 Cas& 30-70 & 5000-7200     & 1.80-2.50     & 2.0-12.5       \\
FU Dra  & 60-90 & 5000-6500     & 7.00-8.50     & 0.1-8.5        \\
UV Lyn  & 50-90 & 5000-6500     & 5.40-7.50     & 2.5-7.0        \\
BB Peg  & 60-90 & 4500-6600     & 5.00-8.00     & 2.0-8.0        \\
V592 Per& 80-90 & 5800-6800     & 2.20-2.90     & 4.0-8.0        \\
OU Ser  & 40-70 & 5000-7000     & 2.06-2.50     & 5.0-12.0       \\
EQ Tau  & 60-90 & 5000-6000     & 2.10-3.30     & 5.5-12.5       \\
HN UMa  & 40-60 & 5700-6500     & 2.00-2.20     & 5.0-13.5       \\
HT Vir  & 50-90 & 5000-6500     & 3.30-6.00     & 3.0-12.5       \\
\hline
\end{tabular}
\label{range}
\end{center}
\label{Tab3}
\end{table}

{\it 2.7. ~OU Ser} \\

OU Ser (BD+16$^{\circ}$2773, HIP 075269, V=8.25$^{m}$) belongs to the W UMa stars 
discovered by the Hipparcos mission. 
It also belongs to the group of high-velocity stars (V$_{\rm sp}$=124.0$\pm$4.4 km/s).
The large proper motion of OU Ser had been the reason for its inclusion in the 
survey by Carney et al. (1994) who noted also the broad lines indicating a 
short-period and the possibility of light variations. 
Pribulla and Va\v{n}(2002) published the first ground-based BV
photometry of the system. They found that the second maximum was fainter
than the first one. This effect was more pronounced in the B filter. Yesilyaprak
(2002) observed   OU Ser in the V filter and determined several times of 
minima. Olsen (1994) obtained Stromgren photometry of the star and derived 
the spectral type of OU Ser to be G1/G2V.
Rucinski et al. (2000) classified the star as F9/G0 and obtained the
following spectroscopic elements of the orbit: $\gamma$=-64 km/s;
K$_{1}$=40.6 km/s; K$_{2}$=234.2 km/s;
q=0.173; (M$_{1}$+M$_{2}$)sin$^{3}$i=0.64 M$_{\odot}$.\\

{\it 2.8. ~EQ Tau} \\

The variability of EQ Tau (GSC~01260-00909, V=10.45$^{m}$) was discovered 
by Tsesevitch (1954). New photometric observations were gathered by 
Benbow and Mutel (1995) and Buckner et al. (1998).
The  light curves of EQ Tau published by Yang and Liu (2002) 
showed  a typical O'Connell  effect.  They also claimed  a variable period 
for this system and classified it as a contact binary of sub-type A. The asymmetry
of the light curve was  explained by  a cool spot.
The most recent photometry in UBV filters was obtained by 
Va\v{n}ko et al. (2004).
Due to low brightness, EQ Tau was not included as a target in the Hipparcos
mission. In the Tycho 2 catalogue it appears with V=11.3 mag and
B-V=0.98.
Qian and Ma (2001) found that the orbital period of EQ Tau was
decreasing and explained it by the mass transfer between components. 
On the basis of the fact that its parameters do not agree with the mass-radius 
relation, they suspected that EQ Tau has a near contact confguration. 
Rucinski et al. (2001) obtained the following parameters of the
spectroscopic orbit: $\gamma$=72 km/s; K$_{1}$=112.4 km/s;
K$_{2}$=254.4 km/s; q=m$_{c}$/m$_{h}$=0.442;
(M$_{1}$+M$_{2}$)sin$^{3}$i=1.75 M$_{\odot}$.\\

\begin{figure}
\begin{center}
\includegraphics[width=8cm,height=12cm,scale=1.0,angle=270]{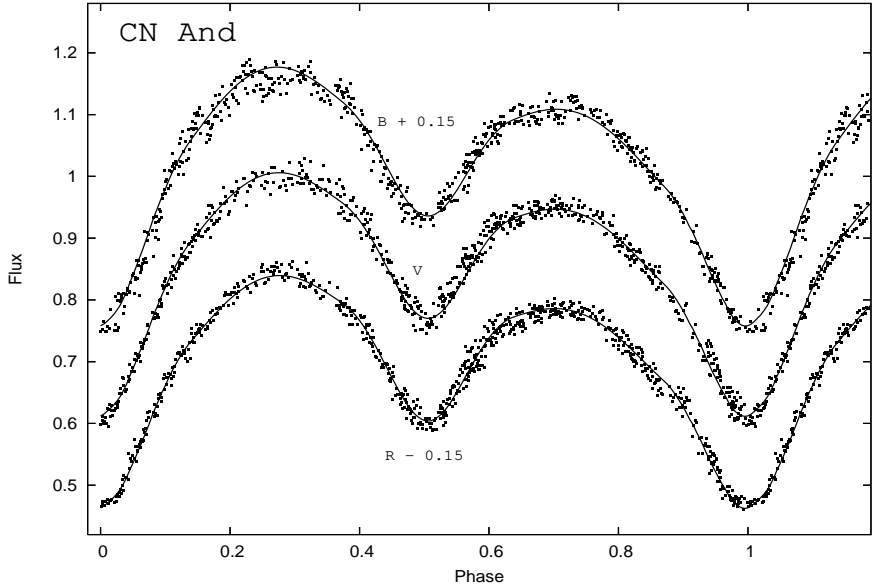}
\caption{Comparison between theoretical and observed light curves
of CN And. Individual observations are shown by dots while lines
represent the theoretical light curves.}
\end{center}
\label{Fig1}
\end{figure}

{\it 2.9. ~HN UMa} \\

HN UMa (BD+38$^{\circ}$2220, HIP 055030, V=9.82$^{m}$) is another eclipsing binary 
discovered by the Hipparcos mission. The times of minima were determined by 
Dvorak (2005) but no ground-based photometry of the system has been published yet.
Using the Hipparcos photometry Selam (2004) computed q=0.10, the
fill-out factor of 20 $\%$ and inclination i=52.5$^{\circ}$.
The spectroscopic orbit of the star was determined by Rucinski et al. (2003). 
They found the system mass ratio q=0.14 and classified the spectrum as F8V.
The orbital parameters were: $\gamma$=-37.1 km/s; K$_{1}$=29.6 km/s; K$_{2}$=212.2 km/s;
(M$_{1}$+M$_{2}$)sin$^{3}$i=0.562 M$_{\odot}$.\\

\begin{figure}
\begin{center}
\includegraphics[width=8cm,height=12cm,scale=1.0,angle=270]{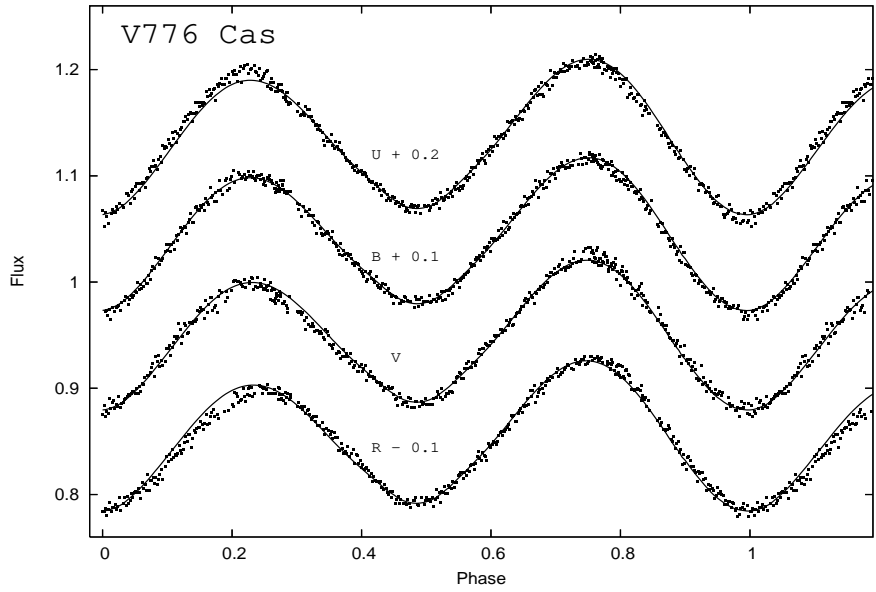}
\caption{Comparison between theoretical and observed light curves
of V776 Cas}
\end{center}
\label{Fig2}
\end{figure}

\begin{figure}
\begin{center}
\includegraphics[width=8cm,height=12cm,scale=1.0,angle=270]{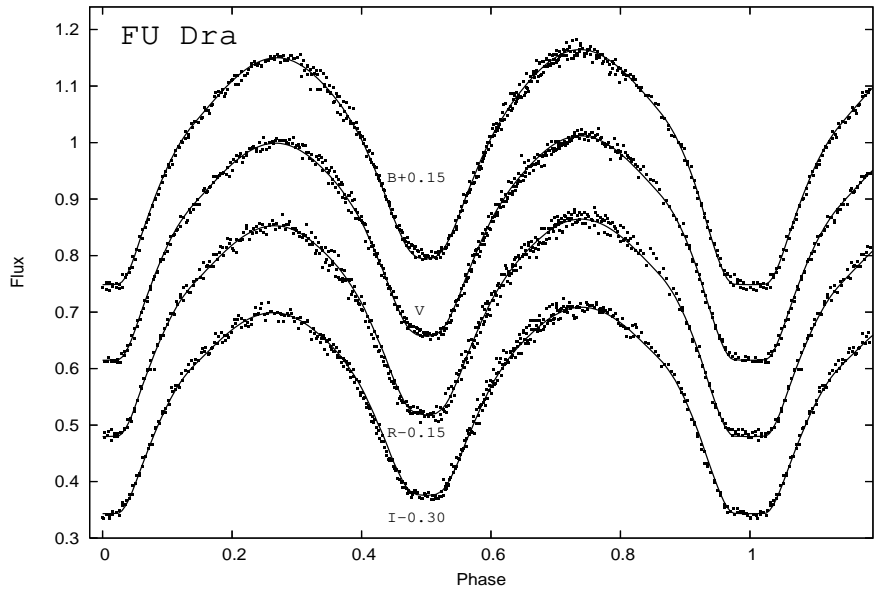}
\caption{Comparison between theoretical and observed light curves
of FU Dra}
\end{center}
\label{Fig3}
\end{figure}

{\it 2.10. ~HT Vir} \\

During observations of the very close visual binary ADS 9019
Walker (1984) discovered variability of HT Vir (BD+5$^{\circ}$2794, HIP67186, 
V=7.162$^{m}$). The light curves
indicated presence of a contact binary of W UMa-type with a period
of 0.4077 days. From analysis of their UBV photometric observations Walker and 
Chamblis (1985) concluded that the components of the eclipsing
pair are identical, they are in contact configuration and the third component 
contributes at the maximum as much light as the eclipsing system itself.
Lu et al. (2001) determined the following spectroscopic elements
for the close binary: $\gamma$=-23.4 km/s; K$_{1}$=169.4 km/s;
K$_{2}$=208.5 km/s; (M$_{1}$+M$_{2}$)sin$^{3}$i=2.285 M$_{\odot}$;
q=0.812. They found that the third component is also a radial
velocity variable with a period of 32.45 days that makes HT Vir 
a quadruple system. Based on the broadening function results Lu et al. (2001)
estimated that the companion(s) were about twice fainter than the contact
binary at its maximum. \\

\begin{figure}
\begin{center}
\includegraphics[width=8cm,height=12cm,scale=1.0,angle=270]{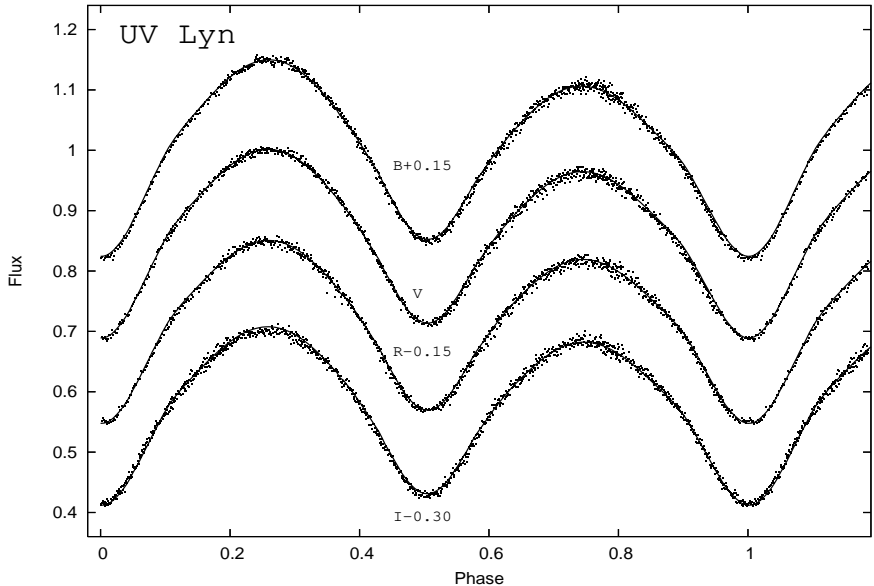}
\caption{Comparison between theoretical and observed light curves
of UV Lyn}
\end{center}
\label{Fig4}
\end{figure}

\begin{figure}
\begin{center}
\includegraphics[width=8cm,height=12cm,scale=1.0,angle=270]{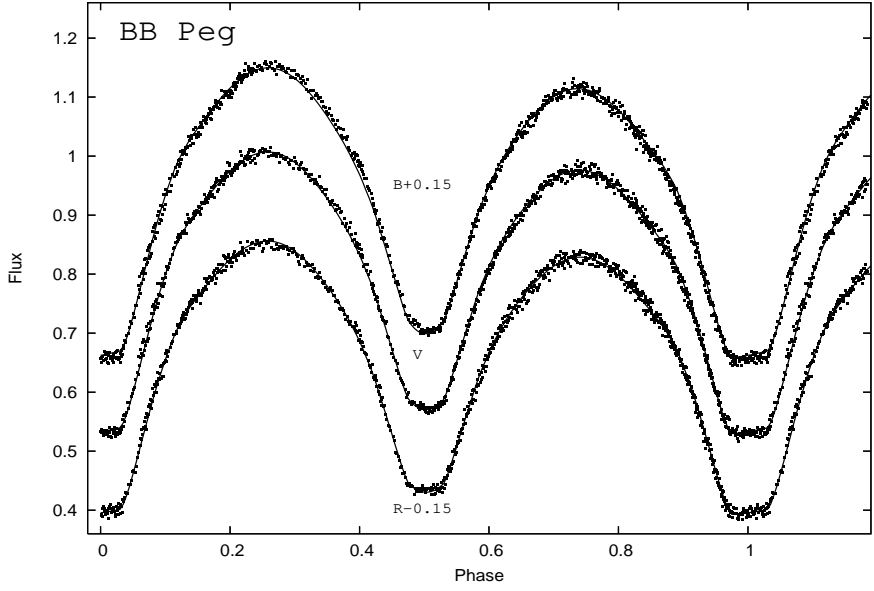}
\caption{Comparison between theoretical and observed light curves
of BB Peg}
\end{center}
\label{Fig5}
\end{figure}

\begin{center}
{\bf 3. ~Photometric observations}
\end{center}

The stars analyzed in this paper were observed at 2 observatories, but we were careful 
to use only one instrument to obtain a complete light curve of each system. 
The light curve of FU Dra was obtained at the Astronomical Observatory of the Jagiellonian 
University with the 50~cm telescope and a Photometrics S300 CCD camera equipped with   
wide-band BVRI filters. We chose GSC~04181-01726 as the comparison star, whose constancy 
was confirmed against GSC~04181-01721.
The photometric data for nine targets  were gathered at Mt. Suhora
Observatory using the two-channel (for CN And only) and  three-channel photometers (for eight 
other systems) attached to the 60-cm Cassegrain telescope. 
Both photometers are equipped with sets of wide-band, close to 
the Johnson-Morgan system filters and Hamamatsu PMTs. UV Lyn and V592 Per were oberved through
the BVRI filters while V776 Cas and HN UMa  through UBVR filters. 
For five targets: CN And, BB Peg, OU Ser, EQ Tau and HT Vir the light curves were 
collected in BVR filters. 
We applied a standard reduction procedure, accounting for dead time, different 
sensitivities of channels, and differential extinction. 
Table 1 \ref{Tab1} presents the journal of observations for objects analyzed in this paper.
The single  data points were phased using the epoch of the primary minimum derived 
from our observations and a recently corrected period (Kreiner 2004).  The ephemerides 
are given in Table 2 \ref{Tab2} along with name of the comparison star we used for 
observations  of each system.

\begin{table}
\caption []{Results from the light curve modeling for CN And, V776
Cas, FU Dra, UV Lyn and BB Peg}
\begin{flushleft}
\begin{tabular}{lrrrrr}
\hline
 parameter                     & CN And            & V776 Cas         & FU Dra            & UV Lyn            & BB Peg             \\
\hline
 configuration                 & near-contact      & contact          &  contact          & contact           & contact            \\
 fill-out factor               &   -               & 77\%             & 15\%              & 18\%              & 21\%               \\
 phase shift                   & 0.0054$\pm$0.0004 &-0.0123$\pm$0.0014&-0.0006$\pm$0.0002 & 0.0033$\pm$0.0002 & 0.0028$\pm$0.0001  \\
 i (degrees)                   & 75.1$\pm$0.1     & 52.5$\pm$0.9     & 81.0$\pm$0.2      & 67.6$\pm$0.1      & 88.5$\pm$0.2       \\
 $T_{\rm 1}({\rm K})$          & *6450             & *6700            & *6100             & *6000             & *6100              \\
 $T_{\rm 2}({\rm K})$          & 5375$\pm$14       & 6725$\pm$90     & 5670$\pm$5       & 5770$\pm$5       & 5780$\pm$7        \\
 $\Omega_{\rm 1}$              & 2.794$\pm$0.013  & 2.001$\pm$0.008  & 7.508$\pm$0.005   & 6.080$\pm$0.001   & 5.936$\pm$0.003    \\
 $\Omega_{\rm 2}$              & 2.619$\pm$0.001  & **2.001          & **7.508           & **6.080           & **5.936            \\
 q$_{\rm corr}$(M$_{\rm 2}$/M$_{\rm 1})$ & *0.371  & *0.138           & *3.756            & *2.685            & *2.589             \\
\hline
 $L^{s}_{1}~(U)$               &                   &  9.355$\pm$0.366 &                   &                   &                    \\
 $L^{s}_{1}~(B)$               & 10.274$\pm$0.039  &  8.993$\pm$0.390 & 3.730$\pm$0.012   & 3.889$\pm$0.011   & 4.249$\pm$0.016    \\
 $L^{s}_{1}~(V)$               & 9.947$\pm$0.037   &  9.224$\pm$0.385 & 3.598$\pm$0.010   & 3.843$\pm$0.009   & 4.194$\pm$0.014    \\
 $L^{s}_{1}~(R)$               & 9.634$\pm$0.035   &  9.520$\pm$0.365 & 3.510$\pm$0.009   & 3.797$\pm$0.008   & 4.116$\pm$0.012    \\
 $L^{s}_{1}~(I)$               &                   &                  & 3.361$\pm$0.007   & 3.772$\pm$0.007   &                    \\
 $L^{s}_{2}~(U)$               &                   & **1.800          &                   &                   &                    \\
 $L^{s}_{2}~(B)$               & **1.977           & **1.738          & **8.501           & **7.766           & **7.586            \\
 $L^{s}_{2}~(V)$               & **2.161           & **1.770          & **8.632           & **7.887           & **7.767            \\
 $L^{s}_{2}~(R)$               & **2.322           & **1.824          & **8.778           & **7.964           & **7.864            \\
 $L^{s}_{2}~(I)$               &                   &                  & **8.891           & **8.116           &                    \\
 l$^{s}_{\rm 3}$~(U)           &                   & 0.080$\pm$0.035  &                   &                   &                    \\
 l$^{s}_{\rm 3}$~(B)           &                   & 0.080$\pm$0.035  &                   &                   &                    \\
 l$^{s}_{\rm 3}$~(V)           &                   & 0.068$\pm$0.037  &                   &                   &                    \\
 l$^{s}_{\rm 3}$~(R)           &                   & 0.047$\pm$0.035  &                   &                   &                    \\
 l$^{s}_{\rm 3}$~(I)           &                   &                  &                   &                   &                    \\
\hline
 $r_{1}~^{side}$        & 0.42975$\pm$0.00026 &0.5947$\pm$0.0036&0.26968$\pm$0.00037&0.2991 $\pm$0.0009 &0.30363$\pm$0.00028 \\
 $r_{2}~^{side}$        & 0.28846$\pm$0.00002 &0.2443$\pm$0.0035&0.50908$\pm$0.00041&0.4797 $\pm$0.0010 &0.47778$\pm$0.00030 \\
\hline \hline
\end{tabular}
\end{flushleft}
 $*$~-~not adjusted,~~~~$**$~-~computed,
~~~~$L^{s}_{1}, L^{s}_{2}$: W-D program input values --
 the subscripts 1 and 2 refer to the star being eclipsed
         at primary and secondary minimum, respectively. \\
\label{Tab4}
\end{table}

\begin{center}
{\bf 4. ~Light curve modeling}
\end{center}

The light curves of all the systems were analyzed by means of the Wilson-Devinney
code (Wilson 1979, 1993) supplemented with the Monte Carlo search method. This 
method does not require any assumption regarding the system configuration, which is concluded 
from the results. The light curve modeling was done simultanously in all filters. 
The procedure we applied was in detail described in Papers I and II. 

\begin{figure}
\begin{center}
\includegraphics[width=8cm,height=12cm,scale=1.0,angle=270]{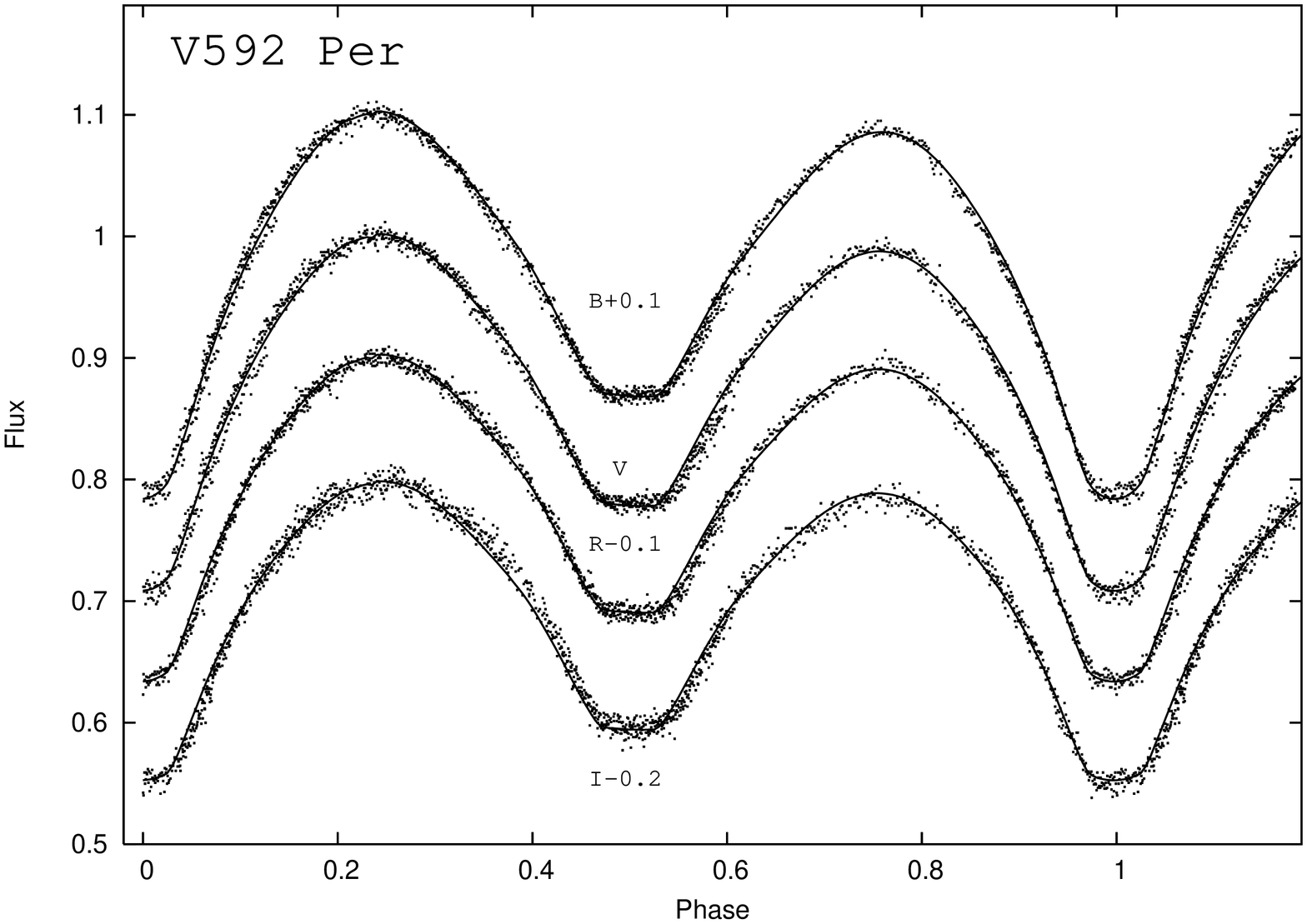}
\caption{Comparison between theoretical and observed light curves
of V592 Per}
\end{center}
\label{Fig6}
\end{figure}

\begin{figure}
\begin{center}
\includegraphics[width=8cm,height=12cm,scale=1.0,angle=270]{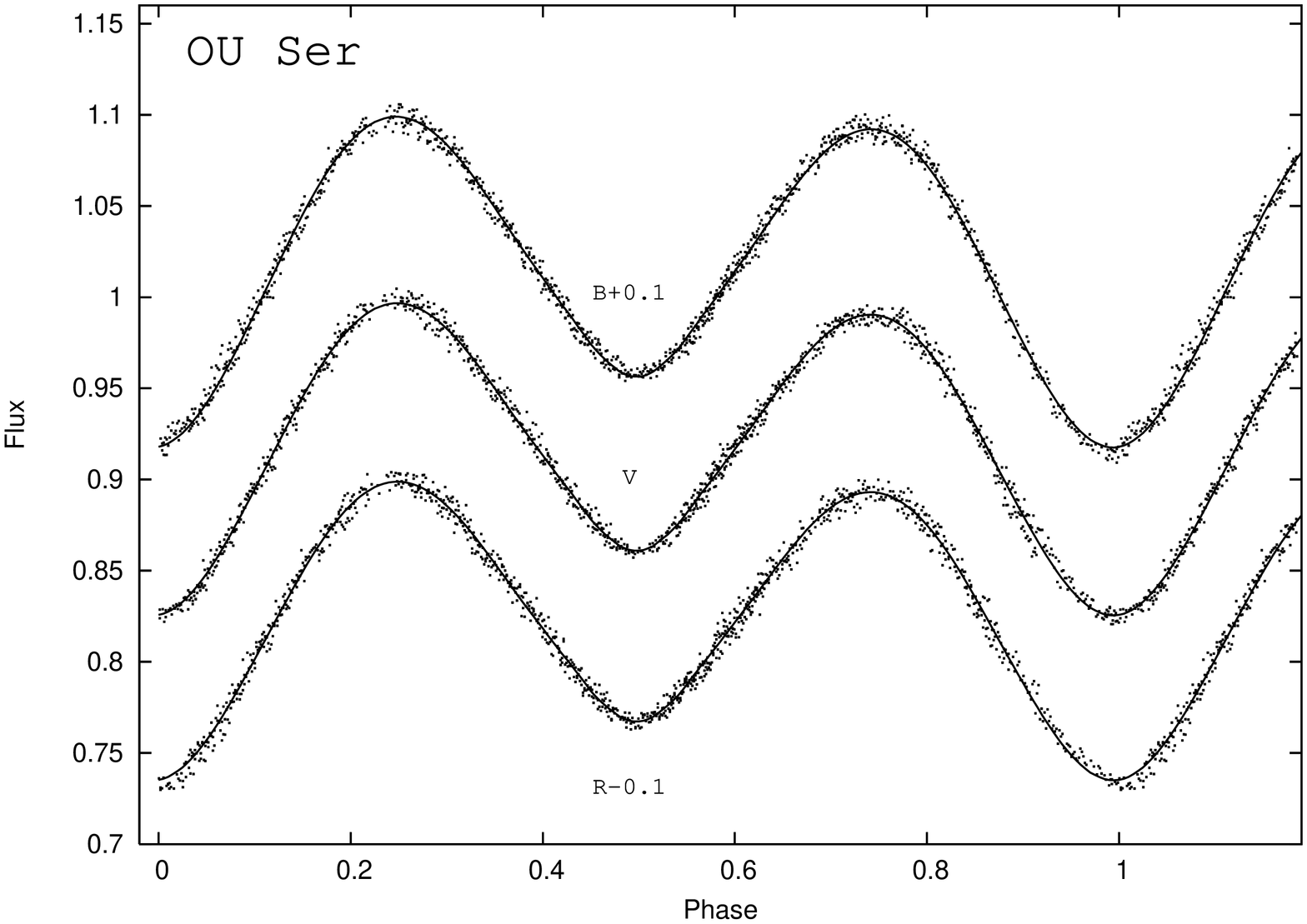}
\caption{Comparison between theoretical and observed light curves
of OU Ser}
\end{center}
\label{Fig7}
\end{figure}

\begin{figure}
\begin{center}
\includegraphics[width=8cm,height=12cm,scale=1.0,angle=270]{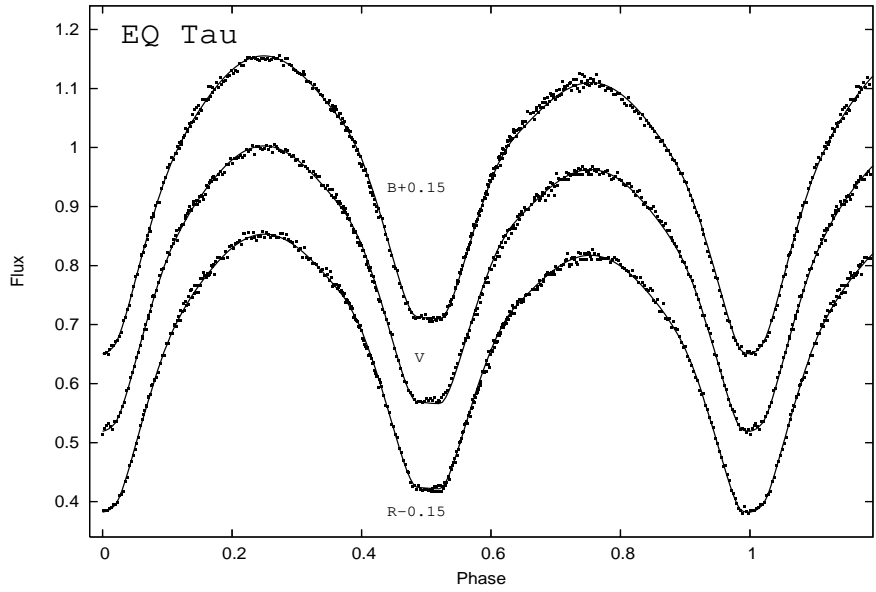}
\caption{Comparison between theoretical and observed light curves
of EQ Tau }
\end{center}
\label{Fig8}
\end{figure}

\begin{figure}
\begin{center}
\includegraphics[width=8cm,height=12cm,scale=1.0,angle=270]{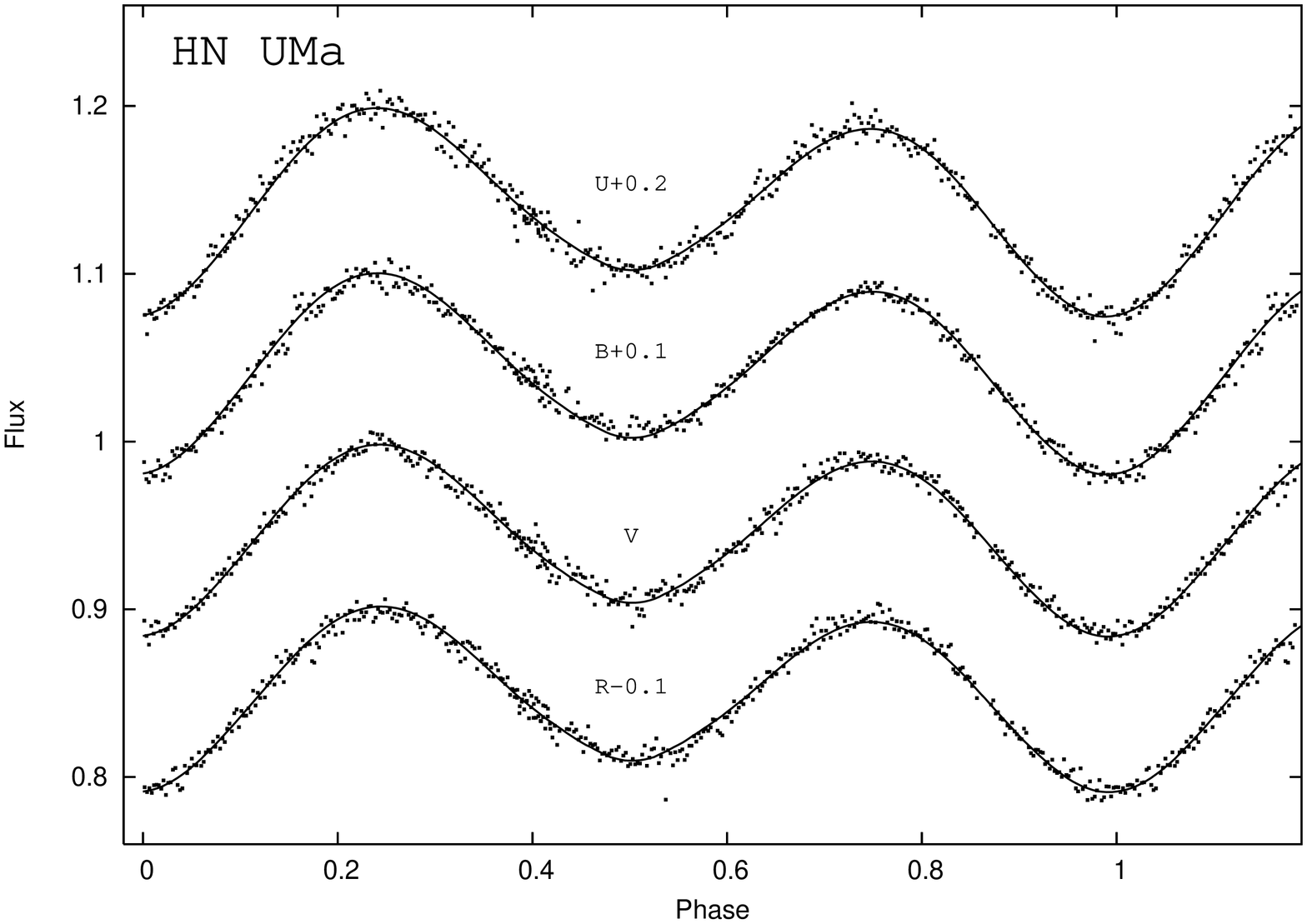}
\caption{Comparison between theoretical and observed light curves
of HN UMa}
\end{center}
\label{Fig9}
\end{figure}

The effective temperature of  star No. 1 (the W-D notation: the star eclipsed at phase 0) 
was fixed and its value was assumed 
according to the spectral type obtained by the DDO group and making use of the 
spectral type versus temperature calibration published by Harmanec (1988). Only the spectral
type of V592 Per was taken from Grenier et al. (1999).
Albedos and gravity darkening coefficients were adopted according
to the theory: A=1.0, g=1.0 for stars with radiative  envelopes 
and A=0.5, g=0.32 for these with convective envelopes. The limb darkening
coefficients were taken from the tables by D\'{\i}az-Cordov\'es et al.~(1995) and 
Claret et al.~(1995). 
The inclination, the temperature of the secondary star, potential(s), phase shift and the 
luminosity of the primary were adjusted. 
In case of an O'Connell effect, clearly seen in the light curve, a spot(s) was added 
to our solution and the whole surface of the brighter component was searched for a 
possible spot location. If there was any information about existence of a companion in
a system also a third light parameter was added and adjusted.
The ranges used for searching for the best solution are given in Table 3.

\begin{figure}
\begin{center}
\includegraphics[width=8cm,height=12cm,scale=1.0,angle=270]{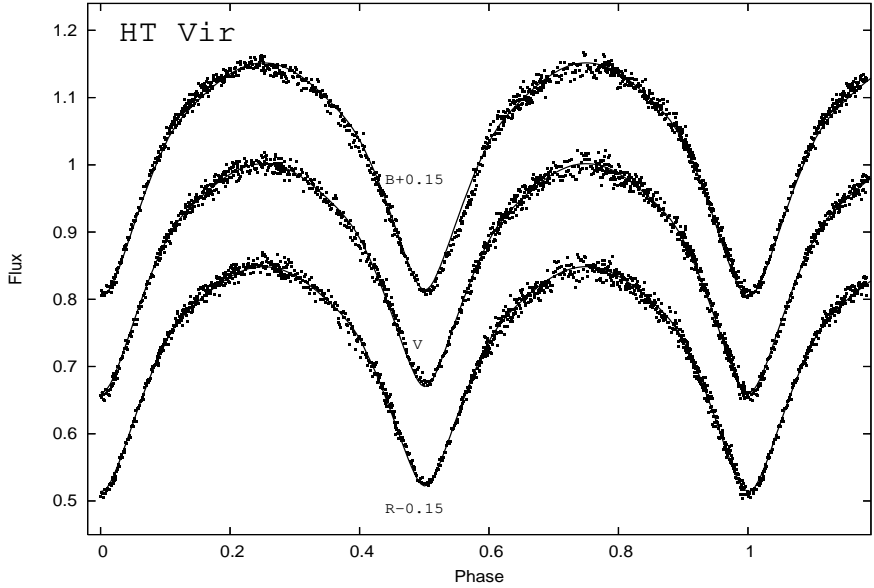}
\caption{Comparison between theoretical and observed light curves
of HT Vir}
\end{center}
\label{Fig10}
\end{figure}

The results from the light curve modeling are presented in  Tables \ref{Tab4} and 
\ref{Tab5}, while the comparison of observations (points) and the best fit achieved 
(lines) for each system is  shown in Figs. \ref{Fig1}1 - 10\ref{Fig10}.\\

\begin{table}
\caption[]{Results from the light curve modeling for V592 Per, OU
Ser, EQ Tau, HN UMa and HT Vir}
\begin{flushleft}
\begin{tabular}{lrrrrr}
\hline
 parameter                         & V592 Per          & OU Ser            & EQ Tau             &  HN UMa            & HT Vir            \\
\hline
 configuration                     &      contact      & contact           &  contact           & contact            & contact       \\
 fill-out factor                   & 59\%              & 44\%              & 13\%               &  32\%              &  8\%              \\
 phase shift                       & 0.0010$\pm$0.0005 & 0.0025$\pm$0.0009 & 0.0029$\pm$0.0001  & 0.0024$\pm$0.0014 & 0.0011$\pm$0.0002 \\
 i (degrees)                       & 89.6$\pm$0.8      & 52.1$\pm$0.8      & 85.5$\pm$0.1       &  46.7$\pm$0.9      & 84.3$\pm$0.5      \\
 $T_{\rm 1}({\rm K})$              & *6800             & *5950             & *5860              &  *6100             & *6100             \\
 $T_{\rm 2}({\rm K})$              & 6020$\pm$40      & 6226$\pm$73      & 5810$\pm$5         &  6082$\pm$77      & 6010$\pm$10      \\
 $\Omega_{\rm 1}$                  & 2.526$\pm$0.008   & 2.113$\pm$0.007   & 2.737$\pm$0.001   &  2.065$\pm$0.007   & 4.067$\pm$0.003   \\
 $\Omega_{\rm 2}$                  & **2.526           & **2.113           & **2.737            &  **2.065           & **4.067           \\
 q$_{\rm corr}$(M$_{\rm 2}$/M$_{\rm 1})$ & *0.389            & *0.172            & *0.447             &  *0.147            & *1.227            \\
\hline
 $L^{s}_{1}~(U)$                   &                   &                   &                    &  9.656$\pm$0.117   &                   \\
 $L^{s}_{1}~(B)$                   & 5.868$\pm$0.073   & 9.087$\pm$0.123   & 8.209$\pm$0.018    &  9.641$\pm$0.119   & 3.553$\pm$0.054   \\
 $L^{s}_{1}~(V)$                   & 5.524$\pm$0.070   & 9.176$\pm$0.112   & 8.192$\pm$0.016    &  9.685$\pm$0.110   & 3.562$\pm$0.053   \\
 $L^{s}_{1}~(R)$                   & 5.162$\pm$0.069   & 9.297$\pm$0.104   & 8.200$\pm$0.015    &  9.772$\pm$0.100   & 3.544$\pm$0.053   \\
 $L^{s}_{1}~(I)$                   & 4.851$\pm$0.073   &                   &                    &                    &                   \\
 $L^{s}_{2}~(U)$                   &                   &                   &                    &  **1.755           &                   \\
 $L^{s}_{2}~(B)$                   & **1.466           & **2.462           & **3.791            &  **1.758           & **3.980           \\
 $L^{s}_{2}~(V)$                   & **1.528           & **2.418           & **3.808            &  **1.770           & **4.026           \\
 $L^{s}_{2}~(R)$                   & **1.523           & **2.377           & **3.829            &  **1.788           & **4.041           \\
 $L^{s}_{2}~(I)$                   & **1.534           &                   &                    &                    &                   \\
 l$^{s}_{\rm 3}$~(B)               & 0.406$\pm$0.005   &                   &                    &                    & 0.362$\pm$0.009   \\
 l$^{s}_{\rm 3}$~(V)               & 0.426$\pm$0.005   &                   &                    &                    & 0.360$\pm$0.009   \\
 l$^{s}_{\rm 3}$~(R)               & 0.457$\pm$0.006   &                   &                    &                    & 0.359$\pm$0.010   \\
 l$^{s}_{\rm 3}$~(I)               & 0.475$\pm$0.007   &                   &                    &                    &                   \\
\hline
 $r_{1}~^{side}$                  &0.4985 $\pm$0.0022 & 0.5624$\pm$0.0028 &0.45900$\pm$0.00009 &  0.5713$\pm$0.0029 &0.3613 $\pm$0.0004 \\
 $r_{2}~^{side}$                  &0.3250 $\pm$0.0021 & 0.2483$\pm$0.0026 &0.31131$\pm$0.00009 &  0.2320$\pm$0.0026 &0.3991 $\pm$0.0004 \\
\hline \hline
\end{tabular}
\end{flushleft}
 $*$~-~not adjusted,~~~~$**$~-~computed,
~~~~$L^{s}_{1}, L^{s}_{2}$: W-D program input values --
 the subscripts 1 and 2 refer to the star being eclipsed
         at primary and secondary minimum, respectively. \\
\label{Tab5}
\end{table}

\begin{center}
{\bf 5. ~Discussion}
\end{center}

The physical parameters for ten close binary systems have been 
obtained by combined solutions of new multicolor light curves and
the radial velocity curves from the DDO spectroscopic program (Table 6).  
The absolute parameters of components of three systems: V592
Per, HN UMa and HT Vir have been determined for the first time.
Our results generally agree with recent determinations based on 
spectroscopic values for the mass ratios for other systems 
(Va\v{n}ko et al. 2001, Djura\v{s}evi\'c et al. 2004, Pribulla and Va\v{n}ko 2002,
Yang and Liu 2002). 

Our new light curves of CN And show a strong asymmetry also noticed by other
observers. The first quadrature  is more than 0.05 mag brighter than the second one
(see Fig. 1). 
In order
to obtain a good fit to the BVR light curves we introduced two spots
in our model and the whole surface of the brighter star was searched for
their location. Due to intrinsic scatter and the distorsion of  the light curve
the exact configuration of this system is not well determined: van Hamme et al. 
(2001) obtained a semidetached configuration with the more massive component filling
its Roche lobe while \c{C}i\c{c}ek et al. (2005) near-contact geometry with 
both components
filling 99\% of their Roche lobes. The final result may even depend on assumptions 
about spots location. Our best solution resulted in one hot and one cool spots
at middle latitudes. The configuration of the best fit is semidetached with
the secondary star filling it Roche lobe. 
The temperature difference between
the components is about 1000~K, as indicated by the unequal depth of the
minima, much higher than these obtained both by van Hamme et al. (2001) and
\c{C}i\c{c}ek et al. (2005). 

The shape of the light curve of V776 Cas indicates low inclination. Additionally
the contribution of a third light further decreases the amplitude of variations.
Our solution resulted in an inclination of i=52.5$^{\circ}$ and the contribution of
the visual companion in the range from 8$\%$ to about 5$\%$ in the U and R filters,
respectively. The secondary star is slightly hotter than the primary. V776 Cas 
is in deep contact configuration with a fill-out factor of 77$\%$. Considering the low
inclination and influence of the third light this solution is not as reliable as
these for high-inlination systems.

Due to its lower brightness observations of FU Dra show larger scatter, but the 
theoretical light curve describes them well. Because of a noticable O'Connel effect, 
not reduced after accounting for color term extinction, one cool spot was 
introduced in our model. We obtained a high inclination of i=81$^{\circ}$
for this system as indicated by deep eclipses. The configuration is contact 
with a  fill-out factor of 15$\%$.

A similar fill-out factor was obtained for UV Lyn (18$\%$). Also this star shows 
a significant difference in brightness of the maxima and we had to add a spot 
into the model. The best fit was obtained for a big cool spot near the equator 
of the secondary star. We confirm the results by Va\v{n}ko et al. (2001) that 
the bigger component in this system is the cooler one.

BB Peg shows a flat bottom primary minimum. The secondary minimum is a little 
shallower. Our solution resulted in a high inclination of i=88.5$^{\circ}$ and a contact
configuration with a fill-out factor of 21$\%$. The bigger component is cooler and
again,  a cool spot was placed on the more luminous
star  as a  noticable O'Connell effect is present in the light curve.

\begin{table}
\begin{center}
\caption[ ]{Absolute parameters of the components}
\begin{tabular}{lcccccc}
\hline
system    & ${\cal M}_{\rm 1}$ & ${\cal M}_{\rm 2}$ & $R_{\rm 1}$ & $R_{\rm 2}$ & $L_{\rm 1}$& $L_{\rm 2}$\\
\hline
CN And    & 1.132$\pm$0.036 & 0.420$\pm$0.023 & 1.252$\pm$0.013 & 0.840$\pm$0.009 & 2.40$\pm$0.05 & 0.52$\pm$0.01\\
V776 Cas  & 1.750$\pm$0.040 & 0.242$\pm$0.017 & 1.821$\pm$0.017 & 0.748$\pm$0.012 & 5.90$\pm$0.11 & 1.01$\pm$0.06\\
FU Dra    & 0.312$\pm$0.012 & 1.173$\pm$0.023 & 0.588$\pm$0.005 & 1.110$\pm$0.009 & 0.42$\pm$0.01 & 1.13$\pm$0.02\\
UV Lyn    & 0.501$\pm$0.015 & 1.344$\pm$0.025 & 0.858$\pm$0.007 & 1.376$\pm$0.010 & 0.84$\pm$0.01 & 1.86$\pm$0.03\\
BB Peg    & 0.550$\pm$0.014 & 1.424$\pm$0.022 & 0.813$\pm$0.007 & 1.279$\pm$0.010 & 0.81$\pm$0.01 & 1.61$\pm$0.03\\
V592 Per  & 1.743$\pm$0.056 & 0.678$\pm$0.037 & 2.252$\pm$0.025 & 1.468$\pm$0.018 & 9.58$\pm$0.21 & 2.50$\pm$0.09\\
OU Ser    & 1.109$\pm$0.038 & 0.192$\pm$0.008 & 1.148$\pm$0.007 & 0.507$\pm$0.006 & 1.48$\pm$0.01 & 0.34$\pm$0.02\\
EQ Tau    & 1.233$\pm$0.030 & 0.551$\pm$0.020 & 1.143$\pm$0.009 & 0.775$\pm$0.006 & 1.36$\pm$0.02 & 0.61$\pm$0.01\\
HN UMa    & 1.279$\pm$0.060 & 0.179$\pm$0.011 & 1.435$\pm$0.010 & 0.583$\pm$0.007 & 2.55$\pm$0.03 & 0.41$\pm$0.03\\
HT Vir    & 1.046$\pm$0.013 & 1.284$\pm$0.015 & 1.107$\pm$0.004 & 1.223$\pm$0.005 & 1.50$\pm$0.01 & 1.72$\pm$0.02\\
\hline
\end{tabular}
\end{center}
\label{Tab6}
\end{table}

V592 Per is another high-inclination system (i=89.6$^{\circ}$). Due to the 
contribution of a companion star, the amplitude variations are smaller than for 
BB Peg. We obtained a deep contact configuration with the contribution 
of the third star to the total light from about 40$\%$ in the B filter to about 48$\%$ 
in the I filter, in good agreement with estimates from broadening functions analysis 
(Rucinski et al. 2005).

Light curves of OU Ser and HN UMa are featurless, similar to that of V776 Cas, indicating
low inclination. Indeed, inclination of 52$^{\circ}$ was derived for OU Ser while even 
lower i=47$^{\circ}$ for HN UMa.
As there is no indication of the existence of a third light in
these systems, their solution should be somewhat more reliable than that of V776 Cas. 
For both systems the contact configuration described observations the best. The 
fill-out factors are 48$\%$ and 32$\%$ for OU Ser and HN UMa, respectively.

The light curve of EQ Tau shows deep eclipses and an O'Connell effect. Therefore,
a cool spot on the primary component surface was assumed. Our solution confirms
the results by Yang and Liu (2002) and Pribulla and Va\v{n}ko (2002). The degree of contact 
is low with the fill-out factor of 13$\%$ and the components have almost the same  
temperatures.

The best fit we obtained for HT Vir resulted in a marginally contact configuration 
with a fill-out factor of 8$\%$. The visual companion contributes about 36$\%$ to 
the total light, independently of the filter, a lower contribution than that derived
by Lu et al. (2001),   Walker and  Chamblis (1985). Since the third light was introduced 
into the model, which is strongly correlated with many parameters, the configuration is 
not quite reliable. Only some independent companion brightness estimate (i.e. from specle 
measurements) can resolve the problem.

\vspace{0.5cm}
{\bf Acknowledgements.} This project was supported by the Polish National
Committee grant No.2 P03D 006 22 and Project No. F1411/2004 of the Bulgarian Ministry
of Education and Science. 
We would like to thank Greg Stachowski for language corrections.\\

\begin{center}
REFERENCES
\end{center}

\noindent
Awadalla, N.S. 1988, Astroph. Space Sci., {\bf 140}, 137.\\
Baran, A., Zola, S., Rucinski, S.M. et al. 2004, Acta Astron., {\bf 54}, 195 (Paper II). \\
Benbow, W., Mutel, R. 1995, IBVS, 4187.\\
Bossen, H. 1973, Astron. Astroph. Suppl. Ser., {\bf 10}, 217.\\
Bozkurt, S., Ibanoglu, C., Gulmen, O., and Gudur N. 1976, IBVS, 1087.\\
Buckner, M., Nellermoe, B., and Mutel, R. 1998, IBVS, 4559.\\
Carney, B., Latham, D., Laird, J., and Aguilar, L. 1994, Astron. J., {\bf 107}, 2240.\\
Cerruti-Sola, M., Milano, L., and Scaltriti F. 1981, Astron. Astroph., {\bf 101}, 273.\\
\c{C}i\c{c}ek, C., Erdem, A., and Soydugan, F. 2005, Astron. Nach., {\bf 326}, 127.\\
Claret, A., D\'{\i}az-Cordov\'es,  J., and Gimenez, A. 1995,  Astron. Astrophys. Suppl. Ser., {\bf 114}, 247. \\
D\'{\i}az-Cordov\'es,  J., Claret, A., and Gimenez, A. 1995,  Astron. Astrophys. Suppl. Ser., {\bf 110}, 329. \\
Djura\v{s}evi\'c, G., Albayrak, B., Selam, S., Erkapi\'c, S., and Senavci, H. 2004, New Astronomy, {\bf 9}, 425.\\
Douglas, G.D., Mason, B.D., Rafferty, T.J., and Holdenried, E.R. 2000, Astron. J., {\bf 119}, 307.\\
Duerbeck, H. 1997, IBVS, 4513.\\
Dvorak, S. 2005, IBVS, 5603.\\
ESA, 1997, The Hipparcos $\&$ Tycho Catalogues, ESA SP-1200, Noordwijk.\\
Evren, S., Ibanoglu, C., Tunca, Z., Akan, M.C., and Keskin, V. 1987, IBVS, 3109.\\
Gazeas, K., Baran, A., Niarchos, P., et al. 2005, Acta Astron., {\bf 55}, 121 (Paper IV).\\
Geier, E., Kippenhahn, R., and Strohmeier, W. 1955, Kleine Veroff. Remeis-Sternw. Bamberg, No 11.\\
Gomez-Forrellad, J., Garcia-Melendo, E., Guarro-Flo, J., Nomen-Torres, J., and Vidal-Sainz, J. 1999, IBVS, 4702.\\
Grenier, S. et al. 1999, Astron. Astrophys. Suppl. Ser., {\bf 137}, 451.\\
Harmanec, P. 1988, Bull. Astron. Inst. Czechosl., {\bf 39}, 329.\\
Heintz, W. 1990, Astroph.J. Suppl., {\bf 74}, 275.\\
Hoffmeister, C. 1931, Astron. Nach., {\bf 242}, 133.\\
Hoffmeister, C. 1949, Astron. Abhandl. Astron. Nach., {\bf 12}, 1.\\
Hog, E., Fabricius, C., Makarov, V., et al. 2000, Astron. Astrophys., {\bf 355}, L27.\\
Hrivnak, B.J. 1990, Bull. American Astron. Soc., {\bf 22}, 129.\\
Kaluzny, J. 1983, Acta Astron., {\bf 33}, 345.\\
Keskin, V. 1989, Astroph. Space Sci., {\bf 153}, 191.\\
Kreiner, J.M., Rucinski, S.M., Zola S., et al. 2003, Astron. Astrophys., {\bf 412}, 465 (Paper I).\\
Kreiner, J.M. 2004, Acta Astron., {\bf 54}, 207.\\
Kuklin, G. 1961, Astron. Tsirk. No 222, 25.\\
Leung, K., Zhai, D., and Zhang, Y. 1985, Astron. J., {\bf 90}, 515.\\
Lu, W.-X, Rucinski, S.M. 1999, Astron. J., {\bf 118}, 515.\\
Lu, W.-X, Rucinski, S.M., and Ogloza, W. 2001, Astron. J., {\bf 122}, 402.\\
Lucy, L., and Wilson, R. 1979, Astrophys. J., {\bf 231}, 502.\\
Markworth, N., and Michaels, E. 1982, PASP, {\bf 94}, 350.\\
Michaels, E., Markworth, N., and Rafert, J. 1984, IBVS, 2474.\\
Olsen, E. 1994, Astron. Astrophys. Suppl. Ser., {\bf 106}, 257.\\
Pribulla, T., and Va\v{n}ko, M. 2002, Contr. Astron. Obs. Skalnate Pleso,  {\bf 32}, 79.\\
Prieur, J.-L., Oblak, E., Lampens, P. et al. 2001, Astron. Astrophys., {\bf 367}, 865.\\
Qian, S.-B., and Ma, Y. 2001, PASP, {\bf 113}, 754.\\
Rafert, J.E., Markworth, N.C., and Michaels, E.J. 1985, PASP, {\bf 97}, 310.\\
Rucinski, S.M., Capobianco, C., Lu, W.-X. et al., 2003, Astron. J., {\bf 125}, 3258.\\
Rucinski, S.M., Lu, W.-X., and Mochnacki, S.W. 2000, Astron. J., {\bf 120}, 1133.\\
Rucinski, S.M., Lu, W.-X., Mochnacki, S.W. Ogloza, W., and Stachowski G. 2001, Astron. J., {\bf 122}, 1974.\\
Rucinski, S.M., Pych, W., Ogloza, W., et al. 2005, Astron. J., {\bf 130}, 767.\\
Samec, R.G., Laird, H., Mutzke, M., and Faulkner, D. 1998, IBVS, 4616.\\
Seeds, M., and Abernethy, D. 1982, PASP, {\bf 94}, 1001.\\
Selam, S. 2004, Astron. Astrophys., {\bf 416}, 1097.\\
Shaw, J.S., Caillault, J., and Schmitt, J. 1996, Astrophys. J., {\bf 461}, 951.\\
Strohmeier, W., Knigge, R., and Ott, H. 1964, Veroff. Remeis-Sternw. Bamberg, {\bf 5}, No 18.\\
Tsesevitch, V.P. 1954, Odessa Izv., {\bf 4}, 110.\\
Van Hamme, W., Samec, R., Gothard, N., Wilson, R., Faulkner, D., and Branly, R. 2001, Astron. J., {\bf 122}, 3436.\\
Va\v{n}ko, M., Parimucha, S., Pribulla, T., and Chochol, D. 2004, Balt. Astr., {\bf 13}, 151.\\
Va\v{n}ko, M., Pribulla, T., Chochol, D. et al. 2001, Contr. Astron. Obs. Skalnate Pleso, {\bf 31}, 129.\\
Walker, R. 1984, IBVS, 2486.\\
Walker, R., and Chamblis, C. 1985, Astron. J., {\bf 90}, 346.\\
Whitney, B. 1959, Astron. J., {\bf 64}, 258.\\
Wilson, R.E. 1979, Astrophys. J., {\bf 234}, 1054.\\
Wilson, R.E. 1993, Documentation of Eclipsing Binary Computer Model.\\
Yang, Y.-L, and Liu, Q.-Y 1985, IBVS, 2705.\\
Yang, Y.-L, and Liu, Q.-Y 2002, Astron. J., {\bf 124}, 3358.\\
Yesilyaprak, C. 2002, IBVS, 5330.\\
Zhang, X.-B, Zhang, R.-X., Zhai, D.-S, and Fang, M. 1995, IBVS, 4240.\\
Zola, S., Rucinski, S.M., Baran, A., et al. 2004, Acta Astron., {\bf 54}, 299 (Paper III).\\

\end{document}